# Asset-Normalized Distribution Reliability Budget Optimization Using Historical Outage Benefits and Work-Type Learning

Catherine Tajmajer
Reliability and Resiliency Planning
Eversource Energy
Berlin, Connecticut
catherine.tajmajer@eversource.com

Hritik Gopal Shah
Reliability and Resiliency Planning
Eversource Energy
Berlin, Connecticut
hritik.shah@eversource.com

***Abstract* - Reliability investments compete for constrained capital but are often evaluated in separate programs. This paper presents a data-driven framework that translates historical work orders, geographic asset records, outage events, and costs into comparable customer-minute and customer-interruption benefits at the circuit-zone level. A normalization process maps historical device names and non-isolating assets to current automatic-isolating-device zones and links investments to pre- and post-construction outage histories. Utility data covering more than $250 million in investments completed from 2010 to 2020 are used to estimate benefit factors for seven work types. Those factors feed a mixed-integer portfolio model that selects at most one applicable action per zone under a capital budget. Observed reductions range from 32% to 94% for customer-minutes interrupted and from 43% to 93% for customer interruptions. The framework provides a traceable path from heterogeneous utility records to zone-level reliability investment decisions.**

## I. INTRODUCTION

Electric distribution companies are simultaneously facing aging infrastructure concerns, increasing climate risk, increased distributed energy resource penetration, electrification-driven load growth, aggressive policy goals, and increasing customer expectations. Traditional distribution investments can materially improve reliability if applied in the correct manner. However, annual planning still commonly begins from siloed program-level budgets: a fixed amount is assigned to distribution automation, vegetation management, cable replacement, or overhead reliability work, and candidate locations are then selected within each program.

That workflow is operationally familiar but economically suboptimal. A planner may know that a reliability investment can reduce outages, but the harder question is: which action is the best next cost-effective decision on the grid? The problem is especially complex because different interventions affect different reliability mechanisms, and actions and outcomes are often interdependent. Reconductoring and tree work often go together and can eliminate certain outage causes. Sectionalizing does not eliminate the fault, but it reduces the number of customers affected and the duration of interruption for some customers. Transformer fusing restricts a potential lateral or backbone outage to a transformer-level outage, reducing customers affected. TripSavers avoid sustained interruptions for faults that can be cleared by de-energizing the conductor but create momentary interruptions. Cable replacement addresses a recurrent underground failure mechanism that typically requires longer repair. To balance the costs and benefits of these myriad applications, a common planning currency is required.

This paper addresses that gap by proposing a practical, utility-scale method for reliability budget optimization. The method differs from traditional feeder-level reliability studies in three ways. First, it evaluates investments at the device or circuit-zone level, where the impact of discrete measures can be segmented and quantified. Second, it converts each observed intervention into empirical customer-minute and customer-interruption savings, enabling cross-program comparison. Third, it carries the learned benefit-cost factors forward into a candidate selection model that ranks all active zones by the most cost-effective applicable mitigation.

The novelty is not merely the optimization formulation. Prior studies have optimized reliability investments. The contribution here is a practical data-to-decision architecture for utility-scale deployment that considers operational and enterprise-level issues. The framework explicitly handles historical outage management system name changes, work-order point-location ambiguity, circuit reconfiguration, device replacement, and outage attribution to non-automatic devices. Without these normalization steps, pre- and post-benefit calculations would miss historical outages, double count events, or associate benefits with the wrong device.

## II. BACKGROUND AND LITERATURE CONTEXT

IEEE Standard 1366 defines standard distribution reliability indices and the factors affecting their calculation [1]. For customer population N, where e denotes an outage reporting record, the common sustained-interruption metrics are defined as:

$$\mathrm{SAIDI} = \frac{\sum_e \mathrm{CMI}_e}{N} \quad (1) \qquad \mathrm{SAIFI} = \frac{\sum_e \mathrm{CI}_e}{N} \quad (2)$$

IEEE Standard 1366 also defines a major event day classification using a log-normal beta method. If d indexes nonzero daily System Average Interruption Duration Index observations over the historical window, then

$$T_{\mathrm{MED}} = \exp(\alpha + 2.5\beta) \quad (3)$$

where alpha is the mean of the logarithm of daily System Average Interruption Duration Index and beta is its standard deviation. Days with daily values exceeding the major-event threshold are removed from normal-operation reliability statistics. This paper focuses on blue-sky and minor-storm reliability improvements because these are the conditions most directly affected by distribution capital reliability programs.

Prior work has established distribution reliability evaluation methods [2], [3] and optimization approaches for reliability investments and protective-device placement [4]–[10]. Recent research also addresses vegetation maintenance and resilience-oriented FLISR methods [11]–[13].

The proposed framework differs from these studies by making utility data transformation part of the optimization process rather than assuming historical events are already associated with a stable network model. This emphasis on data integration and benefit-cost analysis is consistent with broader grid-modernization planning guidance [14]–[18].

## III. DATA ARCHITECTURE AND ASSET NORMALIZATION

### A. Data Requirements

For each investment, the framework requires project population, project cost, program driver, construction work type, asset location, completion date, outage history before and after construction, and benefit calculation rules. In the case study, records came from work management, geographic information, outage management, and financial systems, and engineer-created program tracking files. The source systems changed over time. Detailed device-level outage records were available only after an outage management system transition, and work orders from the modern work management system were most suitable for automated point-location matching. For this reason, the investment-level analysis used work completed between 2010 and 2020, allowing at least four years of post-construction outage history.

A circuit zone is defined as the conductor segments and customers downstream of an automatic isolating device, such as a breaker, recloser, or fuse, to the next automatic isolating device or end of line. Outages within a circuit zone should roll up to the isolating device at the head of the zone. This definition is essential because, in practice, many outages roll up to devices or assets that are not the true head of the circuit zone. Examples include conductors, jumpers, manual solid-blade cutouts or switches. The proposed normalization assigns each event to the correct upstream automatic isolating device used for current planning analysis.

### B. Historical Event Translation

A major implementation challenge is that the same physical protection device can appear under different names over time. For example, a location may be recorded as an old outage-management-system fuse, a single-phase recloser, and later a TripSaver after nomenclature changes. Events may also be assigned to conductors, jumpers, solid blades, or manual switches inside a zone. The consequence is that if only the current device name is queried, historical outages would be missing; if all historical names are queried without normalization, benefits may be double-counted or assigned to the wrong zone. The framework therefore retains original attributes while adding current asset and current-zone identifiers. Without normalization, a pre- and post-analysis might falsely conclude that the current device had no prior outages.

Let $e$ denote an outage reporting record, $a_e^{\text{hist}}$ the original event device or asset, and $G_t$ the best available topology evidence at time t. The normalization process maps the event to a current planning zone:

$$z_e = \Phi\left(a_e^{\text{hist}}, G_t, G_0\right) \quad (4)$$

TABLE I
MAPPING OF EQUATION (4) VARIABLES TO DATA SOURCES

| Symbol | Meaning | Data source |
|---|---|---|
| $a_e^{\text{hist}}$ | Original event device or asset | Outage management records |
| $G_t$ | Event-time topology and alias evidence | Historical geographic/topology records |
| $G_0$ | Current topology snapshot | Current geographic information system |
| $z_e$ | Current planning-zone assignment | Derived normalization output |
| Phi | Identity, isolation, and upstream roll-up rules | Data-integration logic |

Here $G_0$ is the current topology snapshot, and Phi contains three steps: identify the current asset, determine whether it is automatic-isolating, and, if not, identify the upstream automatic-isolating device. Inactive assets are resolved by exact-location match, moved-one-pole logic, and closest-device logic. Manual devices are rolled upstream using geographic device-isolation extracts. This procedure does not merely use a historical asset name to index the current topology: $G_t$ supplies the best available event-time evidence, while $G_0$ supplies a consistent present-day planning frame. Fig. 1 shows the event-device-to-zone-device normalization workflow used in the study.

Because investments were evaluated over approximately a 10-year period, the pre- and post-construction windows spanned years with different weather severity. Outage metrics were adjusted by a dimensionless annual system factor: the corresponding annual system-level customer-minutes or customer interruptions divided by its study-period mean. Equivalently, each zone metric is multiplied by the study-period system mean and divided by the system total for that year. The adjusted metric M therefore retains customer-minute or customer-interruption units, so the cost ratios in (7) and (10) remain dollars per saved unit. This annual adjustment reduces systemwide year-to-year variation but does not isolate the causal effect of investment i from other investments, reconfiguration, or operating changes in the same period.

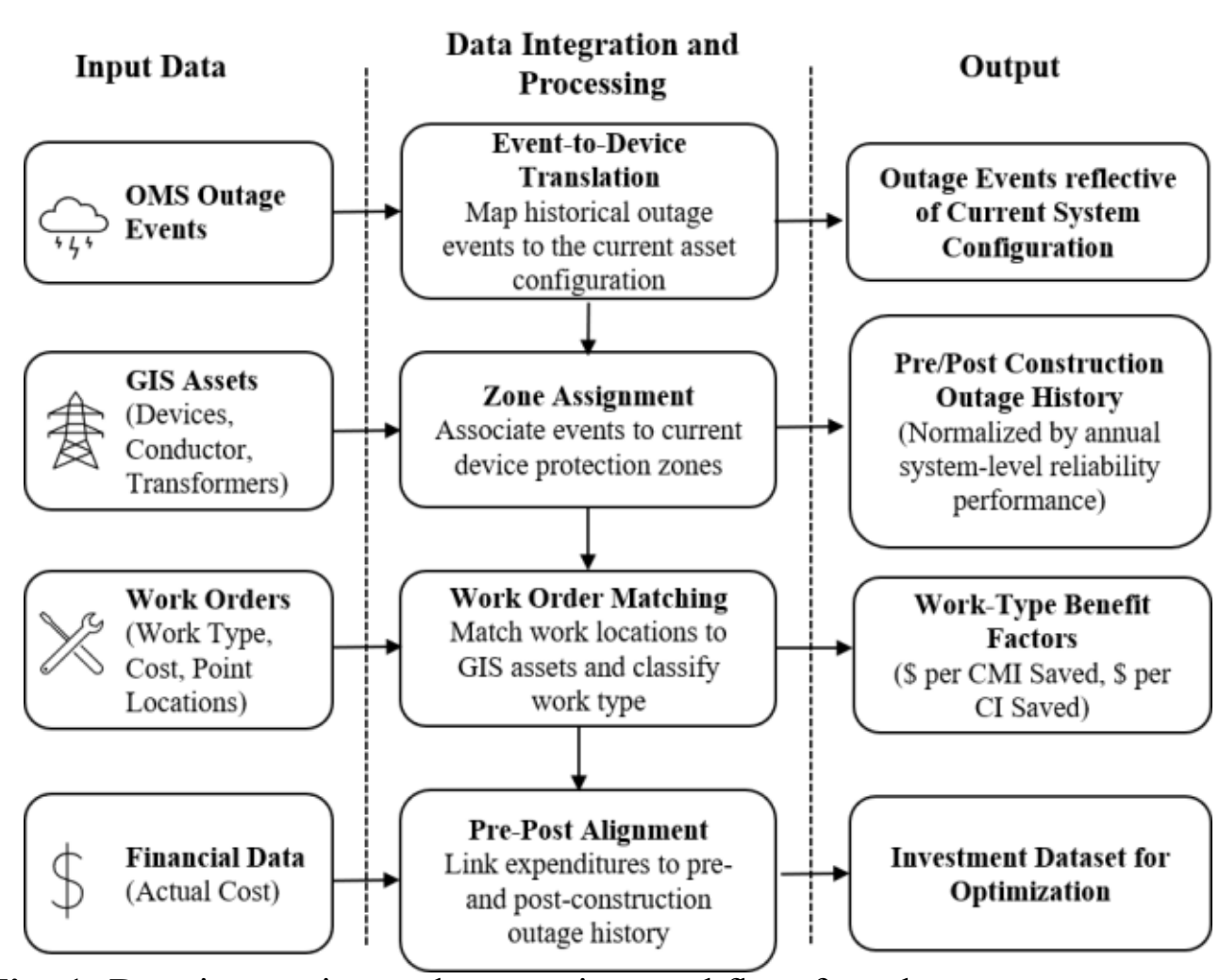

**Fig. 1.** Data integration and processing workflow from heterogeneous source systems to a zone-level investment dataset.

## IV. BENEFIT LEARNING FROM HISTORICAL WORK TYPES

### *A. Pre/Post Estimators*

For each investment $i$, use the average value of metric $m$ (CMI or CI) before and after the construction year $t_i$. Let $M_{i,m,t}$ represent the annual weather-adjusted outage impact associated with investment $i$, measured using metric $m$ in year $t$ The construction year is excluded because it may contain both pre- and post-construction operating conditions:

$$M_{i,m}^{\text{pre}} = \text{avg}_{t<t_i} M_{i,m,t}, \qquad M_{i,m}^{\text{post}} = \text{avg}_{t>t_i} M_{i,m,t} \tag{5}$$

The investment savings are:

$$S_{i,m} = M_{i,m}^{\text{pre}} - M_{i,m}^{\text{post}} \tag{6}$$

For work type $k$, the learned improvement rate and cost per saved metric are computed on totals:

$$r_{k,m} = \frac{\sum_{i\in I_k} S_{i,m}}{\sum_{i\in I_k} M_{i,m}^{\text{pre}}}, \qquad \eta_{k,m} = \frac{\sum_{i\in I_k} C_i}{\sum_{i\in I_k} S_{i,m}} \tag{7}$$

Here $I_k$ is the set of investments of work type $k$ and $C_i$ is actual project cost. The improvement rate $r_{k,m}$ estimates the benefit of a candidate mitigation, and $\eta_{k,m}$ is the historical dollars-per-customer-interruption benchmark. Computing on totals avoids unstable ratios for small or zero-outage projects.

### *B. Work-Type Benefit Logic*

The type of work determines which outage records must be included. Reconductoring and enhanced tree trimming primarily eliminate outage causes; therefore, performance is compared at the upstream isolating zone. Sectionalizing and transformer fusing mitigate outage impact; therefore, the post-construction event set must include the new downstream devices and the original parent zone. For transformer fusing, analyzing each transformer independently would duplicate pre-construction parent events and overstate benefits. For TripSavers, the pre-construction fuse and post-construction TripSaver must be associated as one protection point. For direct-buried cable replacement, the correct spatial unit is often a geographic direct-buried map area rather than an individual underground device.

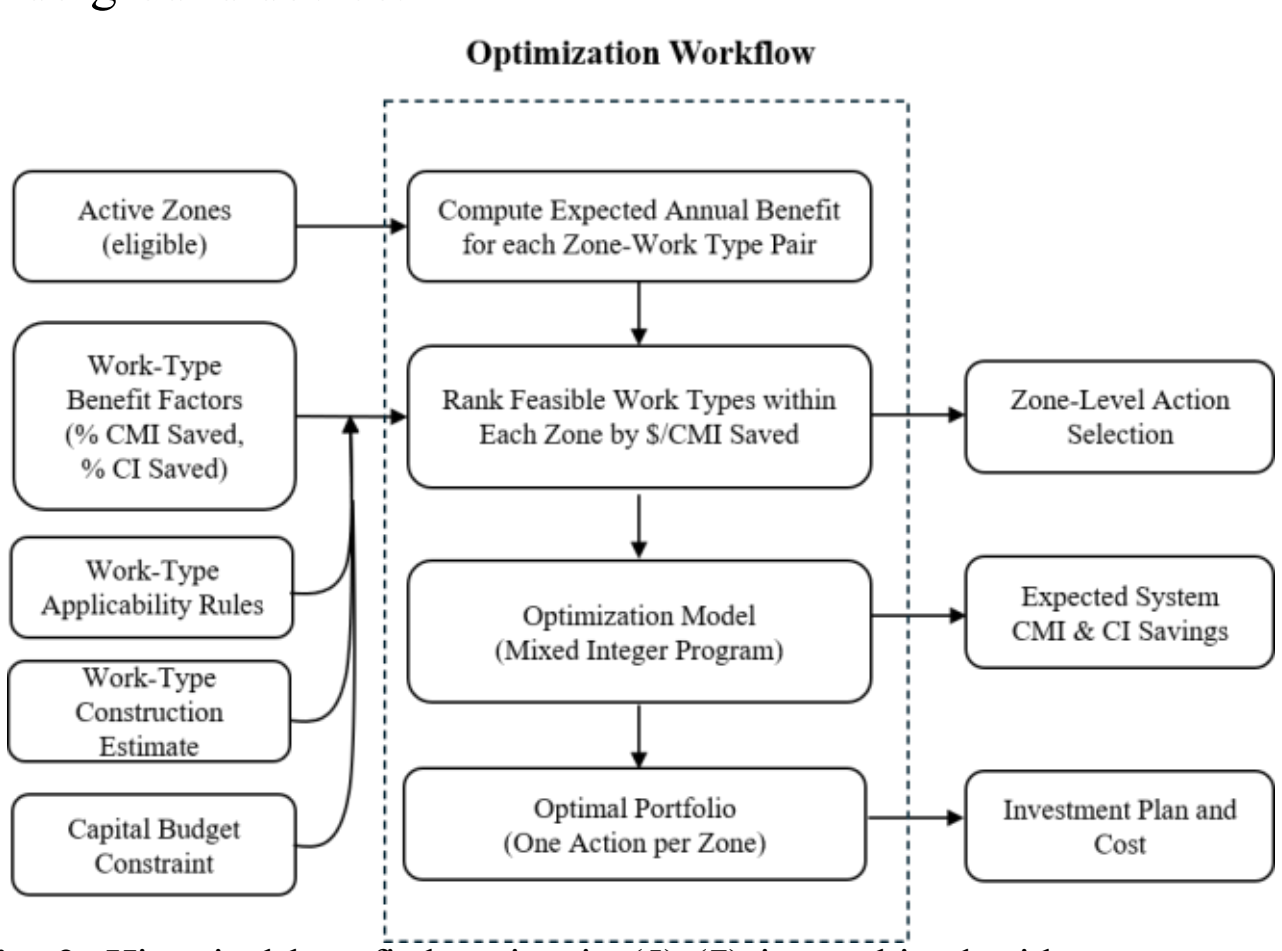

**Fig. 2.** Historical benefit learning in (5)-(7) is combined with current zones and applicability rules to generate candidates in (8)-(11) and select a budget-constrained portfolio using (12).

## V. OPTIMIZATION MODEL

### *A. Candidate Generation*

Let $z$ be a current automatic-isolating-device zone and $k$ a feasible work type. Let $H_{z,m}^{adj}$ be the current lookback value for metric $m$ after the prior-work adjustment defined below. Engineering rules define $a_{z,k} = 1$ if work type $k$ is applicable to zone $z$. Examples include: TripSavers apply to existing fuse locations within fault-current and voltage limits; backbone sectionalizing applies to zones with at least a specified number of customers at risk; reconductoring requires sufficient bare-conductor exposure; enhanced vegetation work applies to zones with tree-related history.

The expected annual benefit uses the same work-type improvement rate $r_{k,m}$ from (7):

$$B_{z,k,m} = a_{z,k}\, r_{k,m}\, H_{z,m}^{\text{adj}} \tag{8}$$

The candidate cost is:

$$C_{z,k} = F_k + u_k q_{z,k} \tag{9}$$

In (9), $F_k$ is the fixed design, mobilization, and setup cost for work type $k$; $u_k$ is its unit cost; and $q_{z,k}$ is the estimated scope quantity in zone $z$, such as conductor length or number of devices. The zone-level ranking signal is:

$$\rho_{z,k,m} = \frac{C_{z,k}}{B_{z,k,m}} \tag{10}$$

where $\rho_{z,k,m}$ is the candidate cost per expected customer-minute or customer-interruption saved. For a single-zone decision, select:

$$k_z^* = \underset{k:\, a_{z,k}=1}{\arg\min}\, \rho_{z,k,m} \tag{11}$$

### *B. Portfolio Selection*

For portfolio planning, $x_{z,k}$= 1 if work type $k$ is selected for zone $z$. The model maximizes weighted benefit subject to the capital budget and one action per zone:

$$\max_x \sum_{z,k,m} w_m\, B_{z,k,m} x_{z,k} \tag{12a}$$

$$\text{s.t.} \sum_{z,k} C_{z,k}\, x_{z,k} \le B_{\max}, \qquad \sum_k x_{z,k} \le 1 \;\; \forall z \tag{12b}$$

$$x_{z,k} \le a_{z,k}, \qquad x_{z,k} \in \{0,1\} \tag{12c}$$

Equation (12a) is the weighted-benefit objective. Constraint (12b) enforces capital budget $B_{\max}$ and permits at most one work type in each zone. Constraint (12c) permits selection only when a work type is applicable to the zone and defines the binary decision domain. The reported customer-minute run uses $w_{CMI}$= 1 and $w_{CI}$ = 0; a separate customer-interruption run reverses those weights. A customer-focused run can add Customer Experiencing Multiple Interruptions after a commensurate scale is specified. The implemented decision-support workflow can solve (12) as a mixed-integer program or approximate it by ranking eligible zone-work pairs. The ranking approach is transparent and easy for field engineers to review; the mixed-integer program provides a formal benchmark when interactions are added.

### *C. Accounting for Prior Improvements*

A zone may rank highly because the four-year lookback contains outages that occurred before a recent mitigation. Rather than exclude such zones entirely, historical outage data can be adjusted to account for known prior work. The model therefore adjusts the lookback metric before applying (8). For a prior work type h already installed in zone z:

$$H_{z,m}^{\text{adj}} = H_{z,m}\big(1 - r_{h,m}\big) \tag{13}$$

Here $r_{h,m}$ is the learned improvement rate from (7) for prior work type h and metric m. If no prior mitigation exists, $H_{z,m}^{\text{adj}}$= $H_{z,m}$. This adjustment prevents the model from taking credit twice for the same historical improvement while still allowing additional work if the adjusted zone remains high-ranked.

## VI. CASE STUDY RESULTS

### *A. Population and Work Types*

The utility data set included over $250 million of reliability-related investments completed from 2010 to 2020. Seven work types were analyzed in the mature results set: 1) direct-buried cable rejuvenation, 2) direct-buried cable replacement, 3) enhanced tree trimming, 4) reconductoring, 5) backbone sectionalizing, 6) TripSaver installations, and 7) transformer fusing. These were selected because they are traditional, reactive measures that are connected to measurable outage histories and customer impact. Proactive asset-health programs, substation programs, and load growth projects were excluded from the optimization training set because their value is not fully visible in pre- and post-blue-sky outage data. Table II summarizes the empirical work-type improvement factors produced by the historical benefit-learning process in (5)-(7). For each work type, the reported CMI and CI changes represent the aggregate observed reduction between the normalized pre- and post-construction periods. These values form the historical benefit factors used by the optimization framework to estimate the expected benefit of applying the corresponding work type to a current candidate zone. Thus, Table II provides the quantitative link between the historical investment analysis shown in Fig. 2 and the candidate benefit calculation in (8).

TABLE II
WORK-TYPE RELIABILITY RESULTS FROM HISTORICAL PRE/POST ANALYSIS

| Work type | Projects | CMI change | CI change |
|---|---|---|---|
| Direct Buried cable rejuvenation | 60 | -94% | -93% |
| Direct Buried cable replacement | 436 | -79% | -70% |
| Enhanced tree trimming | 1,552 | -55% | -59% |
| Reconductoring | 145 | -61% | -43% |
| Backbone sectionalizing | 45 | -53% | -43% |
| TripSaver installations | 516 | -32% | -50% |
| Transformer fusing | 2,669 | -84% | -85% |

The results demonstrate why work types should not be represented by a single generic reliability benefit. Observed CMI reductions ranged from 32% for TripSaver installations to 94% for direct-buried cable rejuvenation, while CI reductions ranged from 43% to 93%. Moreover, the relative effect of a work type differs depending on the reliability objective. For example, enhanced tree trimming and backbone sectionalizing produced similar CMI reductions of 55% and 53%, respectively, but their CI reductions were 59% and 43%. TripSaver installations showed the opposite pattern, with a 32% reduction in CMI compared with a 50% reduction in CI. Consequently, the candidate benefit calculated in (8), and ultimately the portfolio selected in (12), can change depending on whether CMI or CI is used as the optimization objective. Table II reports technical effectiveness only; candidate cost is introduced separately through (9) and the cost-effectiveness calculation in (10).

The data also reveal diminishing returns. In one TripSaver subset, earlier installations produced lower dollars per customer-minute and dollars per customer interruption saved than later years, consistent with the intuition that the first devices are installed on the worst-performing zones and later deployments address less severe locations. This finding further supports dynamic zone-level ranking rather than fixed program budgets, because the marginal value of a work type changes as the highest-value opportunities are addressed.

### *B. Result Interpretation by Work Type*

A few examples illustrate how the results should be interpreted. Sectionalizing reduces customer exposure rather than preventing faults, so it is strongest for high-customer-minute backbone zones. TripSavers are most useful for transient faults, such as tree, animal, or lightning events, that can be cleared by de-energization; earlier deployments were more cost-effective, indicating diminishing marginal returns. Transformer fusing moves animal or transient faults from the parent zone to the transformer device, so parent and transformer events must be analyzed together. Reconductoring, enhanced tree trimming, and direct-buried cable work often reduce failure causes directly, but their costs and asset-health drivers can be higher, so they must be interpreted with the work-type context.

### *C. Optimization Sensitivity*

The same budget can produce different portfolios depending on whether the objective is customer-minutes, customer interruptions, or a customer-focused metric such as Customer Experiencing Multiple Interruptions. A customer-minute objective tends to select high-duration or high-customer-minute zones. A customer-interruption objective favors work that reduces interruption frequency, such as TripSavers and transformer fusing. A repeated-interruption objective can move investment toward repeatedly interrupted laterals even when their contribution to the system duration index is minimal. This is a strength of the proposed model: the historical benefit-cost layer is constant, while the objective weights in (12) can be adjusted for utility, regulatory, or equity priorities. Table III provides a numerical example of how the historical benefit factors are translated into a zone-level decision. For a representative zone, the framework first determines which work types are applicable and the portion of the zone's historical CMI that each action can reasonably address. The applicable CMI is then multiplied by the corresponding historical improvement factor learned from the pre/post analysis to estimate expected CMI savings, consistent with (8). Candidate cost is estimated using (9), and dividing that cost by the expected CMI savings produces the ranking metric in (10).

TABLE III
EXAMPLE ZONE-LEVEL DECISION LOGIC AND COST PER CMI SAVED

| Candidate work type | ETT | Reconductor | TripSaver | Backbone Sectionalize |
|---|---|---|---|---|
| Outage causes addressed | Tree | All | Tree, animal, unknown | All |
| Applicable CMI | 32,000 | 40,000 | 37,000 | 40,000 |
| Historical improvement | 0.55 | 0.61 | 0.32 | 0.53 |
| Average CMI saved | 17,600 | 24,400 | 11,840 | N/A |
| Estimated cost | $100,000 | $500,000 | $23,000 | N/A |
| $/CMI saved | $5.68 | $20.49 | $1.94 | N/A |

**Selected action: TripSaver; lowest $/CMI saved: $1.94.**

The example demonstrates the distinction between technical effectiveness and economic ranking. Reconductoring produces the largest expected benefit in the example, with 24,400 CMI saved, but its estimated cost of $500,000 results in a cost-effectiveness ratio of $20.49 per CMI saved. Enhanced tree trimming produces 17,600 CMI of expected savings at $5.68 per CMI saved. TripSaver produces lower expected savings of 11,840 CMI, but its substantially lower estimated cost results in the lowest available ratio, $1.94 per CMI saved. It is therefore the zone-level action selected under the CMI-based ranking logic in (10) – (11). Backbone sectionalizing is not technically applicable, therefore not ranked.

This example also illustrates why the framework does not simply select the work type having the largest historical percentage improvement in Table II. Historical effectiveness is only one input to the decision. The selected action depends on the outage exposure that the work can address in the specific zone, the learned work-type improvement factor, applicability constraints, and the estimated cost of implementing the action at that location. Equation (11) identifies the preferred action for an individual zone, while the portfolio formulation in (12) extends this comparison across all eligible zone-work pairs and selects the combination of projects that satisfies the available capital budget.

## VII. CONCLUSION

### *A. Why Asset Normalization Matters*

The most important lesson from the study is that reliability optimization is not only a mathematical problem; it is also a data-lineage problem. A model that assumes clean device histories may fail when applied to real utility records. In this data set, one protection location could require five historical names to reconstruct a single outage history. Circuit reconfigurations, nested restoration records, and device replacements can dilute outages across multiple records. Work-order point locations may be sufficient for construction crews but insufficient for automated reliability analysis. The proposed normalization layer makes optimization credible.

### *B. Limitations and Future Scope*

The proposed framework should not be interpreted as a definitive forecast of the system duration or frequency indices, since outcomes are influenced by factors such as weather, vegetation, asset degradation, and operational conditions, and pre/post analysis might be affected by overlapping interventions. Thus, benefit factors should be treated as subject to change, with future incorporation of uncertainty handling and normalization techniques. Future extensions include integrating stochastic benefit modeling, asset-degradation forecasts, customer interruption valuation, and distributed-resource-aware constraints, while expanding to additional work types such as undergrounding, circuit ties, and non-wires alternatives, enabling evolution toward a dynamic zonal-level decision-support tool. Overall, this work presents an asset-normalized reliability budget optimization framework that leverages historical outage data and work-type benefit-cost learning to prioritize zone-level investments under budget constraints, offering a practical, defensible approach for maximizing improvements in customer-minutes, customer interruptions, and customer-centric reliability metrics within utility planning workflows.